\documentclass[10pt]{article}

\usepackage[margin=1in]{geometry}
\usepackage{times}
\usepackage{graphicx}
\usepackage{amsmath, amssymb}
\usepackage{hyperref}
\usepackage{caption}
\usepackage{subcaption}
\usepackage{authblk}
\usepackage{cite}
\usepackage{float}

\title{A Topological Approach to Parameterizing Deep Hedging Networks}

\author{Alok Das, Kiseop Lee}
\affil{Department of Statistics, Purdue University}
\date{} 

\begin{document}

\maketitle

\begin{abstract}
 Deep hedging uses recurrent neural networks to hedge financial products that cannot be fully hedged in incomplete markets. Previous work in this area focuses on minimizing some measure of quadratic hedging error by calculating pathwise gradients, but doing so requires large batch sizes and can make training effective models in a reasonable amount of time challenging. We show that by adding certain topological features, we can reduce batch sizes substantially and make training these models more practically feasible without greatly compromising hedging performance.
\end{abstract}
\vspace{-0.2in}

\section{Introduction}
The classical hedging problem entails replicating the payoff of a contingent claim under a certain stochastic model. While we can find a complete hedging strategy in a complete market like Black-Scholes, a market is in general incomplete, including jump diffusion, and stochastic volatility models. While there are several hedging approaches in an incomplete market, it is often very difficult to get a closed form solution or even calculate numerically. 

Even in a complete market like Black-Scholes, there are drawbacks to this strategy in both execution and the theory it is based on. A traditional asset pricing and hedging method assumes frictionless markets, perfect liquidity, and normally distributed returns among many other conditions. Volatility in between re-balancing periods can also create losses, as our hedging is "too late" to compensate for underlying movements. These losses can accumulate quickly, and since continuous hedging is not feasible, there is not much improvement that can be made. The model itself is also memoryless in the sense that only the current model parameters are relevant; the path of the underlying price process itself is not of concern. 

Thanks to the development of deep neural networks, we can address these difficulties through "Deep Hedging", which involves training neural networks to determine optimal hedging actions without being constrained by assumptions about the market.

Deep hedging \cite{bühler2018deephedging} \cite{françois2025deephedgingoptionsusing} has experienced more development in recent years mainly due to its flexibility in handling market conditions and ability to learn from simulated data. The work itself began when Buehler et al. \cite{bühler2018deephedging} optimized long short-term memory networks to learn global hedging policies by utilizing mean-square error (MSE) and semi-mean-square error (SMSE) as loss functions. Carbonneau \cite{carbonneau2020deephedginglongtermfinancial} built upon this work by considering applications to long-term look-back options while also utilizing non-quadratic loss functions. Imaki et al. \cite{Imaki_2023} also trained networks to hedge look-back options but with the dependency of past hedging actions omitted as a parameter. The work done in \cite{MikkilaKanniainen2022}\cite{2021arXiv210316409C} \cite{neagu2024deephedgingmarketimpact} takes into account factors such as real-world data, transaction costs and market impact. 

The main improvements of this paper come from a broader area of study known as Topological Data Analysis (TDA) which has recently emerged as a way to understand the structure of complex data when viewing it as an embedding in $d$-dimensional space \cite{chazal2021introductiontopologicaldataanalysis}. Rai et al .\cite{rai2024identifyingextremeeventsstock} employed TDA to identify "extreme" events in the stock market via spikes in $L^1$/$L^2$ norms and Wasserstein distances. The work done in terms of training optimization is fairly limited, but recent research has focused on gradient-based approaches \cite{mueller2024fastdeephedgingsecondorder} rather than optimized parameterizations. We do this by incorporating topological features into the model inputs. We evaluate four models: with and without topological features, each trained using batch sizes of 20 and 1000. On an independent set of price paths, the models with topological features achieve lower PnL variance and maintain stable performance even at the smaller batch size.

The paper is organized as follows. We briefly introduce "Deep Hedging" in Chapter 2. Chapter 3 explains our model. We introduce a topological approach in Chapter 4. We provide examples in Chapter 5. Chapter 6 concludes. 

\section{Deep Hedging}
Deep hedging originated in an effort to hedge financial products within incomplete markets by using neural networks to create deterministic hedging policies. As mentioned previously conventional hedging strategies involve many simplifying assumptions and as such are not particularly well suited outside a purely theoretical environment. For practical applications, neural networks allow us to adapt to market conditions rather than ignoring them as a whole. In practice, neural networks utilize gradient descent to find optimal parameters and are particularly useful in convex risk measure environments\cite{maggiolo2025deephedgingnonconvexitylimitations}. By virtue of its construction, neural networks let us specify hedging instruments, cost functions, and any market frictions we want to optimize. This flexibility has been utilized recently in applications to path dependent options like barrier options\cite{chen2023hedging} and bermuda swaptions \cite{oya2024deephedgingbermudanswaptions}, 2 financial products that cannot be completely hedged discretely in an incomplete market. A recurrent neural network structure is common as it not only takes into account the features provided as they evolve over time, but also past hedging decisions. However, over extended time periods, this can be difficult to unravel when back-propagating and can lead to training times that grow exponentially as model depth increases. Mueller et al. \cite{mueller2024fastdeephedgingsecondorder} addressed this issue by adopting a 2nd order optimization scheme in which they precondition gradients by approximating a curvature matrix as block diagonal and Kronecker factored. In doing so, they observed a 75$\%$ improvement in training steps needed to reach a target loss. Our topological approach is more forward-facing and in Chapter 5, we can see the direct impact this has on hedging performance.+

\section{Model}


The underlying spot price and volatility are derived from the Heston Model\cite{heston1993closedform} in which both the underlying and volatility is determined by a stochastic process. The cliquet itself tracks capped cumulative returns over periods of length 20 and is an example of a path dependent option. The payout is updated every 20 steps and gains (losses) in the current time period are not counted until the period has concluded.

\[
\begin{aligned}
dS_t &= \mu S_t \, dt + \sqrt{v_t}\, S_t \, dW_t^S, \\[6pt]
dv_t &= \kappa(\theta - v_t)\, dt + \xi \sqrt{v_t}\, dW_t^v.
\end{aligned}
\]

For our experiments we assign drift $\mu=0.02$, $v_0=0.025$, $\kappa=2.5$, $\theta=0.02$, $\xi=0.6$, with correlation $\rho=-0.5$ for brownian motions $W_t^S$ and $W_t^v$. To define our topological features effectively we define a 3-dimensional Euclidean plane with some $e>0$. We can use the Vietoris-Rips complex (Rips complex) $\mathcal{R}(X, \epsilon)$\cite{chazal2021introductiontopologicaldataanalysis} to capture connectivity between the data points in our moving window. In this setting any subset of $k+1$ points from out data set forms a k-simplex in $\mathcal{R}(X, \epsilon)$ where distances between points is the standard euclidean norm satisfying $||x_i-x_j|| < \epsilon$. By increasing $\epsilon$ we can track when simplexes are formed as more pairwise distances satisfy the distance condition. These updates are referred to as "birth" and "death" rates to signify the creation and destruction of connections between points in $\mathbb{R}^3$. Topologically this process of quantifying connectivity in 3-dimensional space reflects the creation of a persistence diagram for the 0-dimensional homology. For the previously mentioned moving window we calculate the $L^1$ and $L^2$ norms in an effort to measure outliers in our feature space. Intuitively the $L^1$ norm tracks large jumps in the feature set, whereas the $L^2$ norm captures average dependencies over time.

Recurrent neural networks specifically are of great interest here due to its ability to take into account past hedging actions when determining future actions. Since we seek to mimic realistic trading conditions, we can structure our feature set to contain data available up to time $t$. A common choice for a cost function is quadratic hedging error \cite{hientzsch2023reinforcementlearningdeepstochastic}. Intuitively, using this cost function reflects a goal of minimizing PnL spread and bias towards any directional hedging. Our parameterization allows for our network to make more confident hedging decisions, so minimizing PnL spread is of more concern than removing any directional bias. 

Our goal now is to create a model that can hedge a cliquet. To start our construction we define a filtered probability space on a finite time horizon. Specifically, we operate on a probability space $(\Omega, \mathcal{F}, \mathbb{P})$ where $(\mathcal{F}_t)_{t\geq0}$ captures market information at time $t$. We also denote anything measurable at time $t$ with a subscript $t$. Our data is generated for 240 time steps with a cliquet cap $\mu=0.035$. The feature set is created at each time $t$ and a moving window of size 15 is used to generate the topological features. Since our moving window starts to generate data after 15 time steps, we pad our feature set with zeros to remove any inconsistencies within the sizes of components of $\mathcal{F}$ and as such choose to ignore the notational difficulties that would arise as a way of clearly defining that. 
The core feature set itself consists of the spot price $S_t$, realized volatility $v_t$ and the cliquet payout $\psi_t$. 
\[
\psi(x, t)=\max\bigg [ \sum^t_{\substack{i =20, 40, ...,t}}
    \min\!\left(\frac{x_i}{x_{i-1}} - 1, \mu\right), 0\bigg ]
\]

\section{Topological Approach}
Our approach to creating the topological features previously mentioned can be described as a norm based aggregation of the core feature set. We have a set of features (price, volatility, payout) and we seek to understand the shape of that data. We take that data, embed it in 3-dimensional space and build connections between data points to understand how spread out they are. We quantify this data through "birth" and "death" rates \cite{edelsbrunner2002topological} which is essentially an iterative process that determines how far apart each data point is.

To demonstrate improvement via the addition of the topological features, we train 4 models. The models themselves consist of pairwise combinations of batch sizes and features within $\{20, 1000\}$ and $\{no \ TDA, \ TDA\}$. The feature set itself consists of the underlying spot price, realized volatility, the payout of the cliquet at time $t$ and the $L^1/L^2$ norms of the previous 3 features in a rolling window size of 15. The first layer of the network processes the feature set and the previous action separately through unique dense layers of dimension 32, then processes them through a $TanH$ activation function. This output is then processed through 4 stacked LSTM cells, each of dimension 32. The output is then sent to one final dense layer with dimension 1 to reflect the individual action of delta hedging. This architecture itself draws inspiration from \cite{mueller2024fastdeephedgingsecondorder} where our changes come in uniform dimensionality across layers and a lack of a mask to filter hedging instruments(reflects the constant availability of hedging through the spot). Intuitively, the idea behind stacking layers in this manner is to allow the network to learn deeper dependencies as training progresses. Our loss function we seek to minimize is a scaled PnL variance calculation where we denote $\Omega(T)$ the PnL of hedging at time $T$ and $\gamma=1000$.
\\
\\
\[loss=
\gamma\mathbb{V}\bigg[\Omega(T)-\psi(x, T) \bigg ]
\]

\section{Examples}
\begin{figure}[H]
\centering
\begin{subfigure}{0.46\textwidth}
    \includegraphics[width=\linewidth]{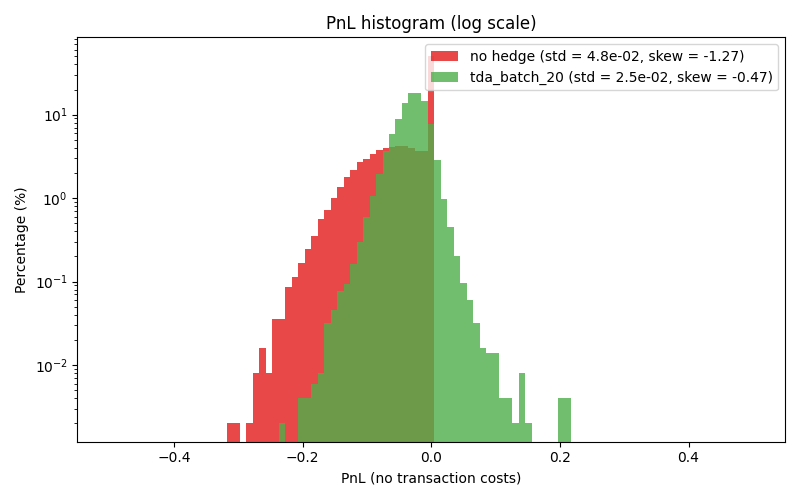}
    \caption{PnL of model using topological features and batch size 20}
    \label{fig:tda_20}
\end{subfigure}
\hfill
\begin{subfigure}{0.46\textwidth}
    \includegraphics[width=\linewidth]{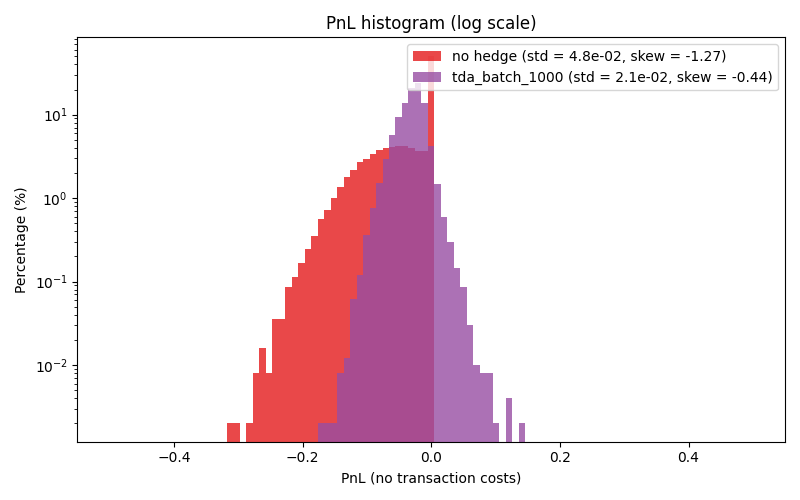}
    \caption{PnL of model using topological features and batch size 1000}
    \label{fig:tda_1000}
\end{subfigure}
\end{figure}

\begin{figure}[H]
\centering
\begin{subfigure}{0.46\textwidth}
    \includegraphics[width=\linewidth]{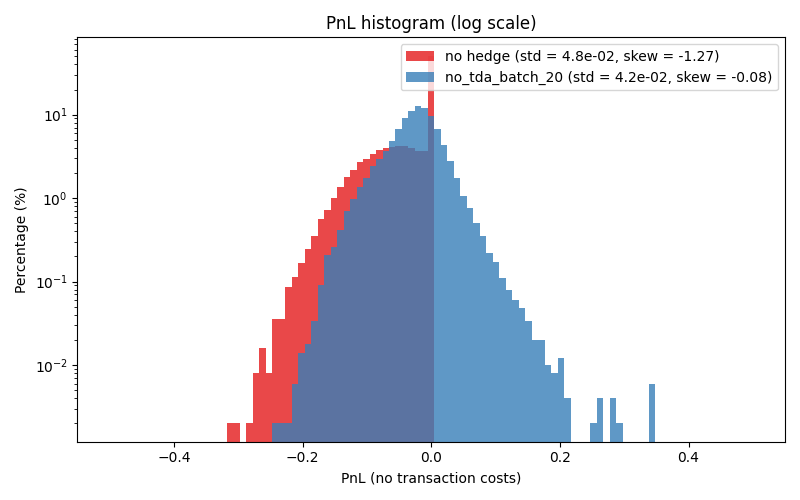}
    \caption{PnL of model without topological features and batch size 20}
    \label{fig:no_tda_20}
\end{subfigure}
\hfill
\begin{subfigure}{0.46\textwidth}
    \includegraphics[width=\linewidth]{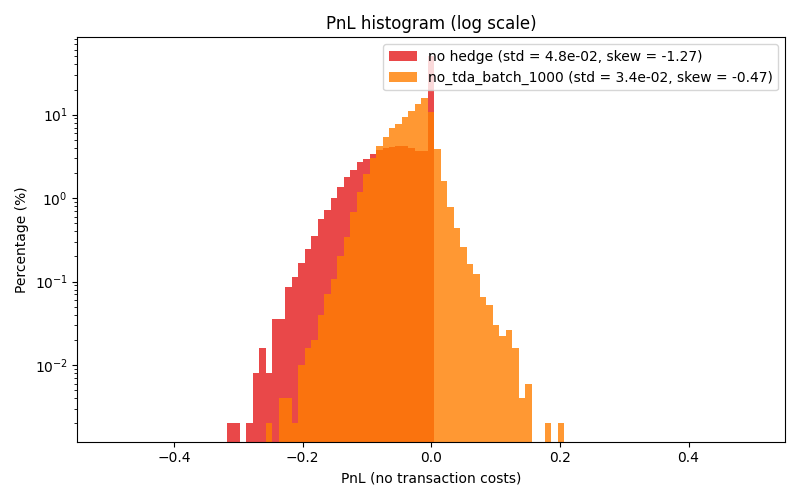}
    \caption{PnL of model without topological features and batch size 1000}
    \label{fig:no_tda_1000}
\end{subfigure}
\end{figure}

Each of the 4 models with their respective batch sizes were trained for $\sim$5300 epochs. Additionally, each of the 4 models was tested on the same 50,000 generated paths to ensure consistency between comparisons in performance. Our metric for a "good" hedging model is removing tail risk and minimizing PnL variance. From Figure~\ref{fig:no_tda_1000}~\ref{fig:tda_1000} we can see that when batch sizes are increased there is a substantial comparative decrease in PnL variance. In both cases when the topological features are added~\ref{fig:tda_20}~\ref{fig:tda_1000} we can see that PnL standard deviation in the batch of 20 goes from 4.2e-02 $\xrightarrow{}$ 2.5e-02 and in the batch of 1000 it goes from 3.4e-02 $\xrightarrow{}$ 2.1e-02. Notably, models that utilize minimal batch sizes tend to have wider distributions. In that specific case the models seem to create more tail probability positive PnL scenarios rather than minimize overall variance.
~\ref{fig:no_tda_20}.
\includegraphics[width=0.9\textwidth]{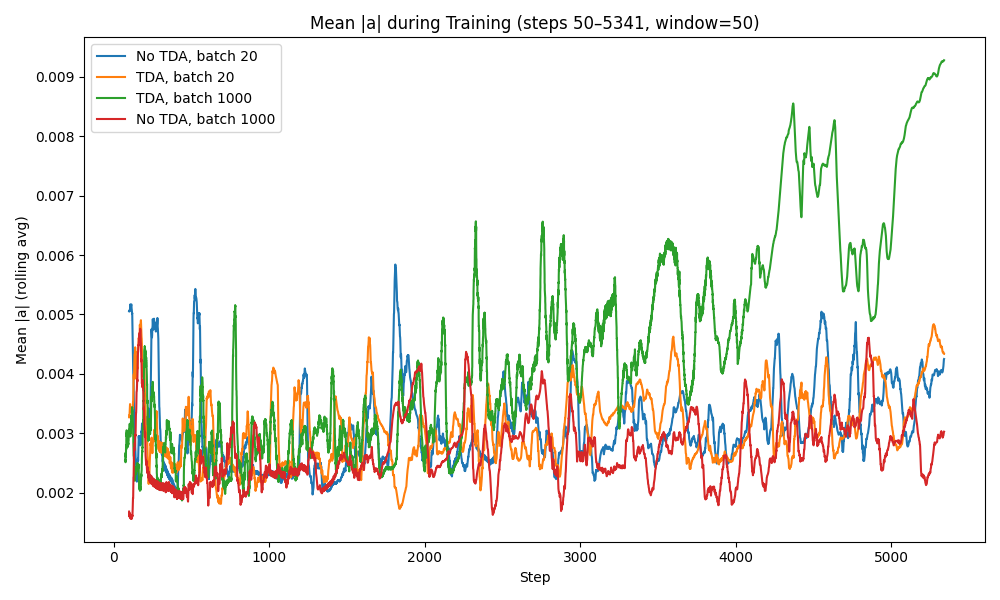}
    \captionof{figure}{Average trade size during training using a rolling average with window size 50}\label{fig:rolling}
    \vspace{2.1ex}
Generally average trade sizes tend to move up as training progresses but for the model containing a large batch size plus the topological features, we start to see a divergence from where the rest of the models seem to cluster. The model containing topological features seemingly learns to trade more as training progresses. In deep hedging applications it is not surprising that a network may learn to trade more actively. As noted in \cite{neagu2024deephedgingmarketimpact}, delta hedging in the presence of limited market liquidity can lead to excessive trading. In our setting, however, we observe that hedging error variance declines despite an increase in average trade size. We interpret this as evidence that the network is not simply overtrading, but is making more decisive hedging decisions. The larger trades reflect greater confidence in its hedging policy, with aggressiveness arising from more accurate assessments of when rebalancing meaningfully reduces risk.

Since the models themselves are all tested on the same training set, we can confidently compare their performances. Our results show that the implementation of topological features not only decreases PnL variance, but allows for smaller batch sizes to achieve similar results. Our data also gives insight into the types of trades being made as well. In chart ~\ref{fig:rolling} we can see that in the case of the 2 models trained with batch sizes of 20, the average trade size is very similar, however in charts ~\ref{fig:tda_20} and ~\ref{fig:no_tda_20} we can see that the topological features lead to a significant decrease in PnL variance. This shows that the network learns to make more meaningful hedges. We can also see in ~\ref{fig:no_tda_1000} that the PnL standard deviation is 3.4e-02 but when adding the topological features ~\ref{fig:tda_1000} the standard deviation drops to 2.1e-02. We can also see the tail-risk PnL outcomes are better under the topological features. Where all PnL outcomes are above -0.2 for ~\ref{fig:tda_1000} and above -0.25 for ~\ref{fig:no_tda_1000}. 

\section{Conclusion}
We proposed a model parameterization that demonstrates comparable hedging performance to models trained with significantly larger batch sizes. Doing so resulted in 10x faster training times. Future work could involve modifying window/batch sizes to find optimal combinations. We experimented with strict rounding masks to simulate a lack of fractional trading availability and did not see much progress, but some improvement to hedging was noted when the topological features were added, so this could also be an avenue of research. Additionally, if we assume batch sizes effect on hedging performance is a convex function it stands to reason that there may exist an optimal batch size besides the ones we tested.

\bibliographystyle{plain} 
\bibliography{references}
\end{document}